**Title:** An Experimental and Computational Study of a Swirl-Stabilized Premixed Flame

**Authors:**

Ashoke De[1], Graduate Student,
Shengrong Zhu[1], Graduate Student
Sumanta Acharya [*,1,2], Professor,

[1]Mechanical Engineering Department
Louisiana State University, Baton Rouge, LA 70803

[2]Turbine Innovation and Energy Research Center
Louisiana State University, Baton Rouge, LA 70803

[*]Corresponding Author

Tel.: +1 225 578 5809 Fax: +1 225 578 5924

E-mail address: acharya@me.lsu.edu

**Keywords**:

Large Eddy Simulation, Thickened-Flame, Swirl, Premixed combustion, Gas turbine, Un-confined configuration

**Abstract**

An unconfined strongly swirled flow is investigated for different Reynolds numbers using particle image velocimetry (PIV) and Large Eddy Simulation (LES) with a Thickened Flame (TF) model. Both reacting and non-reacting flow results are presented. In the LES-TF approach, the flame front is resolved on the computational grid through artificial thickening and the individual species transport equations are directly solved with the reaction rates specified using Arrhenius chemistry. Good agreement is found when comparing predictions with the experimental data. Also the predicted RMS fluctuations exhibit a double-peak profile with one peak in the burnt and the other in the un-burnt region. The measured and predicted heat release distributions are in qualitative agreement with each other and exhibit the highest values along the inner edge of the shear layer. The precessing vortex core (PVC) is clearly observed in both the non-reacting and reacting cases. However, it appears more axially-elongated for the reacting cases and the oscillations in the PVC are damped with reactions.

**Nomenclature**

| | |
|---|---|
| A | pre-exponential constant |
| $C_s$ | LES model coefficient |
| $D_i$ | molecular diffusivity |
| E | efficiency function |
| $E_a$ | activation energy |
| $S_{ij}$ | mean strain rate tensor |
| $T_a$ | activation temperature |
| U | mean axial velocity |
| $U_o$ | bulk inlet velocity |
| $u_i$ | velocity vector |
| $u'$ | rms turbulence velocity |
| W | mean tangential velocity |
| w' | tangential RMS velocity |
| $x_i$ | Cartesian coordinate vector |
| $Y_i$ | species mass fraction |

**Greek symbols**

| | |
|---|---|
| $\Delta$ | mesh spacing |
| $\nu_t$ | kinematic turbulent eddy viscosity |
| $\bar{\rho}$ | mean density |
| $\omega_i$ | reaction rate |

**INTRODUCTION**

Land-based gas turbines operate primarily in a lean premixed mode (LPM) with natural gas as the fuel of choice due, in part, to environmental regulations of reducing NOx. Swirl is used to provide flame-holding, and plays an important role in premixed gas turbine combustors [1, 2]. Several experimental studies have been reported that characterize the flame structure and provide insight into the flame-turbulence interaction in laboratory scale burners [3-5]. Since experiments are generally expensive to undertake, in order to properly design premixed combustion systems, accurate predictions of premixed flames are desirable. The capability of the classical approach using Reynolds averaged Navier-Stokes (RANS) equations in conjunction with phenomenological combustion models [6] is limited from an accuracy viewpoint. Therefore, numerical simulations of reacting flows based on large eddy simulations (LES) have been proposed for providing accurate and cost-effective predictions. The main philosophy behind LES of a reacting flow is to explicitly simulate the large scales of the flow and reactions, and to model the small scales. Hence, it is capable of capturing the unsteady phenomenon more accurately. The unresolved small scales or sub-grid scales must be modeled accurately to include the interaction between the turbulent scales. Since the typical premixed flame thickness is smaller than the computational grid ($\Delta$), the small scale or sub-grid scale modeling must also take care of the interaction between turbulence and the combustion processes.

In turbulent premixed combustion, a popular approach is to rely on the flamelet concept, which essentially assumes the reaction layer thickness to be smaller than the smallest turbulence scales. The two most popular model based on this concept are the flame surface density model (FSD) [7] and the G-equation model [8, 9]. It has been reported that the FSD model is not adequate beyond the corrugated flamelet regime [10-11], while the G-equation approach depends on a calculated signed-distance function that represents an inherent drawback of this method.

Another family of models relies on the probability density function (PDF) approach [12]. This is a stochastic method, which directly considers the probability distribution of the relevant stochastic quantities in a turbulent reacting flow. The PDF description of turbulent reacting flow has certain theoretical benefits; the complex chemistry is taken care of without applying any *ad-hoc* assumptions (like 'flamelet' or 'fast reaction'). Moreover, it can be applied to non-premixed, premixed, and partially premixed flames without having much difficulty. However, the major drawback of the PDF transport approach is its high dimensionality, which essentially makes the implementation of this approach to different numerical techniques, like FVM (Finite Volume Method) or FEM (Finite Element Method), limited.

In this work, a Thickened Flame (TF) model [13] is invoked where the flame is artificially thickened to resolve it on computational grid points where reaction rates from kinetic models are specified using reduced mechanisms. The influence of turbulence is represented by a parameterized efficiency function. A key advantage of the TF model is that it directly solves the species transport equations and uses the Arrhenius formulation for the evaluation of the reaction rates.

The configuration of interest in the present work is that of a swirl-stabilized flame. An extensive review on swirling flows can be found in [4, 14]. Chanaud [15] reported periodic vortex instabilities in a certain regime of Reynolds number and swirl numbers. These were identified to be the precision vortex core (PVC). Tangirala et al. [16] studied a non-premixed swirl burner where they reported that the mixing and flame stability can be improved with swirl upto a swirl number of about unity, beyond which a further increase in swirl reduces the turbulence level as well as the flame stability. Broda et al [17] and Seo [18] experimentally investigated the combustion dynamics in a lean-premixed swirl stabilized combustor. As the swirl number exceeds a critical value, vortex breakdown takes place and leads to the formation of an internal recirculation zone [19]. The shape and size of the recirculation zone largely depends on swirl and Reynolds numbers [5].This recirculation not only enhances fuel-air mixing, but also carries hot products back to the reactants and plays an important role in the flame holding. However, despite several years of research, the mechanisms of vortex breakdown are only partially understood [14, 20-21].

In this investigation, we employ both computational and experimental methods, to investigate premixed swirl-stabilized flames. The experimental approach uses particle image velocimetry (PIV) and intensified CCD imaging of flame CH-chemiluminescence. The computational method uses LES combined with a TF approach for combustion. A key task is the assessment of LES-TF model predictions through validations with measurements in this study. A second task is to use the validated model predictions to analyze the flow and combustion physics and, in particular, to explore how increasing flow velocities alter the vortical structures and the associated heat release behavior which play an important role in the flame holding and blow-off behavior.

**EXPERIMENTAL CONFIGURATION AND TECHNIQUES**

The configuration considered here is an unconfined swirl burner. The experimental setup includes the combustor, the PIV system for velocity measurement, and the PI-MAX ICCD camera (Princeton Instruments) for CH-emissions measurement. The combustor consists of the inlet fuel and air-delivery system, and the premixing section. The flame is swirl-stabilized and attached to the center body at the dump plane for conditions corresponding to the measurements in this study.

The 45º swirl vane is fitted with a solid center body which also acts as a fuel injector (Fig. 1). This center body extends beyond the swirl vane and is flush with the dump plane of the combustor. The diameter of the center body is 12.7mm (0.5 inch) and the O.D. of the swirler is 34.9 mm (1.375 inch). Methane gas is injected radially from the center body through eight holes immediately downstream of the swirler vane. The fuel/air mixer is assumed to be perfectly premixed at the dump plane and the equivalence ratio is calculated to be $\varphi=0.7$. The geometric swirl number, defined as the ratio of the axial flux of the tangential momentum to the product of axial momentum flux and a characteristic radius, is $S_g=0.82$. Experiments are conducted at atmospheric pressure and temperature for two different Reynolds numbers of 10144, and 13339 (based on inlet bulk velocity and hydraulic diameter).

**Stereoscopic PIV measurements**

For three-dimensional velocity measurements, a commercial PIV system (IDT Inc) using two Sharp Vision 1300DE cameras is used. These CCD cameras have a resolution of 1280(H)

×1024(V) pixels with pixel size of 6.7 ×6.7µm. Both cameras are equipped with a 50mm Nikon lens. To illuminate the flow field of interest, a pulsed Nd:YAG laser light sheet at 532 nm is used. During the measurement, the PIV system is operated at a 10 Hz frame rate. The time between two laser pulses is between 20 and 40 µs, depending on the flow velocity. The field of view (FOV) is approximately 85mm×60mm, and the lens aperture is adjusted to ensure the appropriate image size for the seeding particles to ensure the accuracy of post processing the data. The seeding particles are small enough to ensure good tracking of the fluid motion (low Stokes number) and big enough to scatter light for image capturing. Here $TiO_2$ particles with nominal diameter of 3µm are introduced upstream of the swirler in order to distribute them homogenously and to follow the flow oscillation with a frequency up to 1 kHz [22].

IDT pro-VISION software with adaptive interrogation mode is utilized; it is based on a second-order accurate mesh free algorithm [23], and is designed to reduce errors associated with loss of pairing, image truncation, and spatial averaging of velocity gradients. A 60×50 mesh has been used to get 3000 vectors per frame with 32×32 correlation windows yielding a spatial resolution of approximately 1.1×1.0 mm. Because of the complex nature of the swirling flow field, care is taken to optimize inter-frame timing, camera aperture setting, laser-sheet thickness, and seeding density. Sets of 500 image pairs are usually recorded for each data set and statistically processed for the mean and RMS values.

**NUMERICAL DETAILS**

**Flow modeling using LES**

To model the turbulent flow, LES is used where the energetic larger-scale motions are resolved, and the small scale fluctuations are modeled. Therefore, the equations solved are the filtered governing equations for the conservation of mass, momentum, energy and species transport in a curvilinear coordinate system [24]. The sub-grid stress modeling uses a dynamic Smagorinsky model where the unresolved stresses are related to the resolved velocity via a gradient approximation:

$$u_i u_j - u_i u_j = -2\nu_t \overline{S}_{ij} \tag{1}$$

where
$$\nu_t = C_s^2 (\Delta)^2 |\overline{S}| \tag{2}$$

$$\overline{S}_{ik} = \frac{1}{2}\left( (\vec{a}^m)_k \frac{\partial \overline{u}_i}{\partial \xi_m} + (\vec{a}^m)_k \frac{\partial \overline{u}_k}{\partial \xi_m} \right) \tag{3}$$

$$|\overline{S}| = \sqrt{2 \overline{S}_{ik} \overline{S}_{ik}} \tag{4}$$

and S is the mean rate of strain. The coefficient $C_s$ is evaluated dynamically [24] and locally-averaged.

**Combustion modeling**

Modeling the flame-turbulence interaction in premixed flames requires tracking of the thin flame front on the computational grid. In the present paper we use the thickened flame approach which is a cost-effective strategy while allowing the chemistry to be represented. In this technique, the flame front is artificially thickened to resolve it on the computational grid while allowing the flame to propagate at the same speed as the un-thickened flame [13]. The artificial thickening of the flame front is obtained by multiplying the diffusion term by a factor F and dividing the reaction rates by the same factor to maintain the flame speed. More detailed description of this technique is found in the literature [24].

The major advantage associated with this TF model is the ability to capture the complex swirl stabilized flame behavior which is often found in a gas turbine combustor. Since this type of geometry with the premixing section does not guarantee a perfectly premixed gas at the dump plane, the fully premixed assumption in the numerical model is not valid any more. The present

TF model is capable of taking care of this type of partially premixed gas since we solve for the individual species transport equations and the reaction rates are specified using Arrhenius expressions.

In the LES framework, the spatially filtered species transport equation using artificial thickening is written as

$$\frac{\partial \overline{\rho} Y_i}{\partial t} + \frac{\partial}{\partial x_j}(\overline{\rho} Y_i u_j) = \frac{\partial}{\partial x_j}\left(\overline{\rho} EFD_i \frac{\partial Y_i}{\partial x_j}\right) + \frac{E\dot{\overline{\omega}}_i}{F} \tag{5}$$

where the modified diffusivity ED, before multiplication by F to thicken the flame front, may be decomposed as ED=D(E-1)+D and corresponds to the sum of molecular diffusivity, D, and a turbulent sub-grid scale diffusivity, (E-1)D. In fact, (E-1) D can be regarded as a turbulent diffusivity used to close the unresolved scalar transport term in the filtered equation. Since the thickened flame does not respond to turbulence like the initial flame, the sub-grid scale effects have been incorporated into the thickened flame model, and parameterized using an efficiency function E derived from DNS results [13]. The detailed description of E is found in the literature [24].

**Chemistry model**

As all the species are explicitly resolved on the computational grid, the TF model is best suited to resolve major species. Intermediate radicals with very short time scales can not be resolved. To this end, only simple global chemistry has been used with the thickened flame model.

In the present study, a two step chemistry, which includes six species ($CH_4$, $O_2$, $H_2O$, $CO_2$, $CO$ and $N_2$) is used and given by the following equation set.

$$CH_4 + 1.5 O_2 \rightarrow CO + 2 H_2O \tag{6}$$

$$CO + 0.5 O_2 \leftrightarrow CO_2 \tag{7}$$

The corresponding reaction rate expressions are given by:

$$q_1 = A_1 \exp(-E^1_a/RT)[CH_4]^{a1}[O_2]^{b1} \tag{8}$$

$$q_2(f) = A_2 \exp(-E^2_a/RT)[CO][O_2]^{b2} \tag{9}$$

$$q_2(b) = A_2 \exp(-E^2_a/RT)[CO_2] \tag{10}$$

where the activation energy $E^1_a = 34500$ cal/mol, $E^2_a = 12000$ cal/mol, a1=0.9, b1=1.1, b2=0.5, and $A_1$ and $A_2$ are 2.e+15 and 1.e+9, respectively, as given by Selle et al. [25]. The first reaction (Eq. 6) is irreversible, while the second reaction (Eq. 7) is reversible and leads to an equilibrium between CO and $CO_2$ in the burnt gases. Hence the expression in Eq. 8 represents the reaction rates for the irreversible reaction (Eq. 6) and the expressions Eq. 9 & 10 represent the forward and backward reaction rates for the reversible reaction (Eq. 7). Properties including density of mixtures are calculated using CHEMKIN-II [26] and TRANFIT [27] depending on the local temperature and the composition of the mixtures at 1 atm.

**Solution procedure**

In the present study, a parallel multi-block compressible flow code for an arbitrary number of reacting species, in generalized curvilinear coordinates, is used. Chemical mechanisms and thermodynamic property information of individual species are input in standard Chemkin format. Species equations along with momentum and energy equation are solved implicitly in a fully coupled fashion using a low Mach number preconditioning technique, which is used to effectively rescale the acoustics scale to match that of convective scales [28]. An Euler differencing for the pseudo time derivative and second order backward 3-point differencing for physical time derivatives are used. A second order low diffusion flux-splitting algorithm is used for convective terms [29]. However, the viscous terms are discretized using second order central differences. An incomplete Lower-Upper (ILU) matrix decomposition solver is used. Domain decomposition and load balancing are accomplished using a family of programs for partitioning unstructured graphs and hypergraphs and computing fill-reducing orderings of sparse matrices, METIS. The message communication in distributed computing environments is achieved using Message Passing Interface, MPI. The multi-block structured curvilinear grids presented in this paper are generated using commercial grid generation software GridPro$^{TM}$.

**Computational domain and boundary conditions**

As noted earlier, and shown in Fig. 2, the configuration of interest in the present work is an unconfined swirled burner. The computational domain extends 20D downstream of the dump plane (fuel-air nozzle exit), 13D upstream of the dump plane (location of the swirl vane in Fig. 2) and 6D in the radial direction. Here, D is the center-body diameter. Two different LES grids are studied (for cold flow only): one that consists of 210x138x32 grid points downstream of the dump plane plus (64x23x32)+(75x17x32) grid points upstream (where the grid is in two blocks), and corresponds to approximately 1.22M grid points (mesh1: coarse). The finer mesh consists of 320x208x48 grid points downstream of the dump plane plus (98x32x48)+(114x22x48) grid points upstream, and contains approximately 3.94M grid points (mesh2: fine).

The inflow boundary condition is assigned at the experimental location immediately downstream of the swirler blades. The mean axial velocity distribution is specified as a one-seventh power law profile to represent the fully developed turbulent pipe flow, with superimposed fluctuations at 10% intensity levels (generated using Gaussian distribution). A constant tangential velocity component is specified as determined from the swirl vane angle. Convective boundary conditions [30] are prescribed at the outflow boundary, and stress-free conditions are applied on the lateral boundary in order to allow the entrainment of fluid into domain. The time step used for the computation is dt=1.0e-3.

**RESULTS AND DISCUSSION**

We will first present the measurements and predictions for the non-reacting LES calculations to ensure that the grid and boundary conditions are properly chosen, and to assess the cold-flow flow characteristics. This will be followed by a discussion of the reacting flow calculations where we will examine both the flow and heat release distributions.

**Non-reacting flow results**

Figure 2 shows the stream line patterns for the two Reynolds number. Three distinct recirculation regions are observed in the high Reynolds number case, Re=13339, that include a separation wake recirculation zone (WRZ) behind the center body, a corner recirculation zone (CRZ) due to sudden expansion of combustor configuration, and a central toroidal recirculation zone (CTRZ) formed due to vortex breakdown. The CTRZ, however, is not prominent in the low Reynolds number case (Re=10144) and appears as an asymmetric structure that becomes more clearly visible and symmetric at higher Reynolds number (Re=13339). The asymmetry at the lower Reynolds number indicates a low-frequency unsteadiness (Fig. 7) that is not averaged out despite the long integration times (15-25 flow through times) used for statistical averaging. Thus the origins of the CTRZ at the lower Re appear to be in the form of a flapping vortical structure that becomes more steady and well defined at higher Reynolds numbers. Based on these observations, at the lower Re, the WRZ and CRZ are likely to play an important role in the flame holding, while with increasing Re, the CTRZ becomes the dominant structure and will be of primary significance in the flame holding behavior.

The radial distribution of the axial and tangential mean velocity profiles and the axial and tangential fluctuations at different axial locations are shown in Fig. 3 for Re=13339. The time-averaged mean quantities are normalized by the corresponding bulk velocity ($U_o$=9.57 corresponds to Reynolds number Re=13339). Results from both the coarse and fine grids are shown in the plots. In general, the agreement between LES and the experimental data is quite good with the peak velocities and turbulence levels correctly predicted both in magnitude and location. The shape, size, and the intensity of the recirculation zone (region of negative axial velocities at the center) are well predicted along with the overall spreading of the turbulent swirling jet. Some level of asymmetry can be observed in both the simulations and the experiments and indicate that the statistical averaging period needs to be carried out over a longer period of time. However, due to the presence of low-frequency unsteadiness in the flow, the averaging time-periods can be very large and impractical from both computational and experimental perspectives. Similar observations of asymmetry in the averaged profiles have also been reported in the literature for a confined combustor geometry [25].

The RMS fluctuations of the axial and tangential velocities are also shown in Fig. 3. The LES predictions only report the resolved stresses, but these predictions are in excellent agreement with the experimental data. The predicted axial fluctuations indicate that the primary contributor to these stresses is from the larger resolved scales. The peak in the axial velocity fluctuations is observed to be in the shear layer and between the location of the peak velocity and the recirculation bubble. In this region, the steepest velocity gradient $\partial U_i/\partial x_j$ is obtained and promotes the production of the peak kinetic energy. The tangential velocity fluctuations show a flatter profile than the axial velocity fluctuations and their peaks are shifted radially inwards as for the mean tangential velocities.

For both the mean velocity and fluctuations, the fine mesh (3.94M grid points) results are in better agreement with the experimental data compared to those from the coarse mesh (1.22M grid points) for all the cases considered [24]. Hence, the fine mesh is chosen for the reacting flow calculations.

**Reacting flow results**

Comparing to non-reacting cases (Fig. 2), the stream line patterns corresponding to the shear layers are quite different due to the added heat release, and the shear layers appear to be more distinct and axially-directed compared to the non-reacting case as observed in Fig. 4. It can be seen that the heat release distributions dramatically alter the flow patterns. Notably, the length of inner recirculation zone (both the WRZ and CTRZ) is reduced while the length of corner-recirculation zone increased quite significantly in the reacting flow field.

Contours of the mean temperature levels are superimposed on the streamlines in Fig. 4 and show that the highest temperatures occur along the inner edge of the shear layers and in the CTRZ. Note the inner edges of the shear layer are associated with the highest turbulence which are likely to promote molecular-level mixing and combustion.

Figure 5 shows the snapshots of the iso-vorticity surface at $\omega=13$ s$^{-1}$ for both reacting and non-reacting cases. It is clearly observed that for low Reynolds number case, a typical vortex spiral starts evolving from the shear layers due to Kelvin-Helmholtz instabilities in both the axial and azimuthal directions. This structure, called the PVC, precesses around the centerline and sustains for several turns before breaking down into smaller structures. For the higher Reynolds number, the spiral vortex structures are also observed; however, the structures appear to be more complex due to the higher centrifugal force and they spread more rapidly in the radial direction before breaking down to form smaller scale structures.

While the PVC is clearly observed for both the non-reacting cases and reacting cases, the PVC appears more axially-elongated for the reacting cases, but suppressed radially (Fig. 5). To understand this behavior, it is educational to look at the terms of the vorticity transport equation which is written as:

$$\frac{D\vec{\omega}}{Dt} = \underbrace{(\vec{\omega}\cdot\vec{\nabla})\vec{u}}_{I} - \underbrace{\vec{\omega}(\vec{\nabla}\cdot\vec{u})}_{II} - \underbrace{\frac{\vec{\nabla}\rho \times \vec{\nabla}p}{\rho^2}}_{III} + \underbrace{\nu\nabla^2\vec{\omega}}_{IV} \quad (11)$$

where the RHS terms are: (I) Vortex stretching, (II) Gas expansion, (III) Baroclinic production, and (IV) Viscous diffusion. Figure 6 shows these terms at one Reynolds number (13,339) for both the non-reacting and reacting flow fields. The vortex stretching term is comparable for both the reacting cases and the non-reacting cases and is not shown since it is not responsible for the differences seen in Fig. 5. The gas expansion term acts as a sink in reacting cases due to negative sign in the transport equation. This term is directly proportional to the gas dilatation ratio across the flame ($\rho_u/\rho_b$), which increases as the temperature increases in presence of combustion. This term and its greater value for the reacting case (Fig. 6) is clearly partly responsible for the reaction-induced-damping of the PVC seen in Fig. 5. As the kinematic gas viscosity increases due to temperature in the flame, this substantially enhances the rate of vorticity diffusion and further dampens the core vorticity. However, the pressure gradient generated due to inclination and expansion of the flame with respect to the flow field contributes to the baroclinic production of vorticity. The gas expansion ratio ($\rho_u/\rho_b$) affects both the source and sink terms. In the present case, it actually gives rise to the production and gas expansion terms and diffusion term as well (Fig. 6). Hence, the combined effects of gas expansion, production and diffusion terms make the PVC axially elongated. Moreover, the size of the corner recirculation zone (CRZ) in Fig. 4 also supports the existence of the axially elongated PVC, although the effects of this exothermicity produce thicker vortical structures in the reacting cases (Fig. 5).

Figure 7 shows the frequency spectra of the velocity field at a specific location in the PVC. In both cases, the PVC oscillating frequencies are below 12 Hz. For the non-reacting case, the low frequency oscillation peaks are more clearly observed with two dominant peaks around 1.5 and 3 Hz. This clearly indicates that the PVC oscillation frequencies are suppressed with reaction.

Figure 8 shows the distributions of the normalized axial velocity profiles, and axial fluctuations at different axial locations for Re=13339. The overall agreement of the predictions with the data is found to be quite reasonable, considering the complexity of the physical processes and the configuration. Compared to the non-reacting cases, the magnitude of the velocity peak is increased and the radial-spreading angle is wider. With increasing axial distance the magnitude of the peak velocity decreases and the location of the peak is moved further outwards radially. As noted earlier by comparing the streamline plots in Figs. 2 and 4, the negative velocities for the non-reacting cases are larger in magnitude than the reacting flow cases.

While the general agreement between the data and predictions are satisfactory, and the LES results show the right qualitative features and the peak magnitudes, there are intrinsic differences between the predictions and data. The axial velocities show a narrower shear-layer region and a small over-prediction of the peak axial velocity particularly at the upstream locations. Predicted RMS fluctuations clearly exhibit two peaks. The location of the peaks correspond to the burnt and un-burnt regions in the inner part of the shear layer and associated with the high velocity gradients where the turbulence production due to the mean velocity gradient is the highest. The first peak is lower in magnitude and located in the burnt region of the shear layer downstream of the center body where the temperatures are higher. The high temperatures cause the viscosity value to go up, and this reduces the magnitude of the peak stress component. The second peak is observed in the un-burnt regions of the shear layer where the temperatures are relatively on the lower side which reduces the viscosity value and in turn exhibits higher fluctuating components. Similar trends for RMS fluctuations have also reported by other researchers reporting calculations [31]. However, the measurements do not clearly show this dual-peak behavior.

The radial distributions of the major species mass fractions are presented in the Fig. 9 for Re=13339. The entrainment of ambient air is clearly observed in the distributions of $O_2$. Along the centerline, $CH_4$ and $O_2$ consumptions are higher which, in turn, reflect in the higher temperature zones observed earlier (Fig. 4). Un-burnt $CH_4$ is observed behind the center-body. No reacting species excluding $O_2$ is observed away from the flame region (r1>20 mm). Away from the dump plane (X4=40 mm) the temperature goes down; consequently the $O_2$ consumption level goes down. This lower temperature zone is clearly visible from Fig. 4 and Fig. 10 where the predicted mean heat release patterns show shorter flame length and lower temperature regions.

**CH chemiluminescence measurement**

CH radicals are produced at the flame front and represent the reaction zones. Therefore CH chemiluminescence imaging has been carried out in the present work [25], and Fig. 10 shows such an integrated image across the flame (left). CH chemiluminescence is considered here to be representative of the heat release rate as shown in Fig. 10, which also shows the predicted heat release distributions along the center plane (middle) and averaged across the flame (right). It should be noted that the CH data and the predicted heat release distributions while related to each other are not directly the same quantity and can only be qualitatively compared with each other, and show reasonable agreement (compare the averaged measured image on left with the averaged predicted image on the right in Fig. 10).

Figure 10 shows that the CH distributions and the heat release patterns are similar to each other, in that, the levels are highest in the inner-regions of the shear layer. It can be seen that the turbulent flame exhibits its peak CH signal closer to the dump plane. This is seen in both the experimental and computational distributions. It is observed experimentally that the flame is slightly lifted off; however, computations show a more compact flame and shorter flame length (also seen in mean velocity predictions, Fig. 8). While not shown, a higher Re tends to increase the turbulence intensity and the flame speed and consequently shortens the flame length. Further, as Re increases, higher turbulence enhances the flame-turbulence interaction resulting in higher heat release. Turbulent kinetic energy (TKE) is usually lower in the higher temperature regions and increases in the lower temperature regions. Thus, with reference to Fig. 10 (middle) the peak TKE is obtained on either side of the maximum CH release and explains the double peak captured in the predictions of the RMS fluctuations of the axial velocity (Fig. 8).

**CONCLUSIONS**

PIV measurements and LES with a TF model are used to investigate unconfined swirling flows in a laboratory based model combustor. Both reacting and non-reacting flow conditions for different Reynolds numbers are studied. A 2-step chemical scheme is invoked to represent the flame chemistry for methane-air combustion. The equivalence ratio for the flame is 0.7 and the geometric swirl number for the configuration is 0.82. CH chemiluminescence imaging is also carried out to characterize the heat release distributions.

Isothermal flow predictions are in good agreement with the measurements and indicate that the boundary conditions and grid are properly chosen. Reynolds number is seen to have an impact on the flow field particularly for the non-reacting cases. At a high Re, all the recirculation zones, such as WRZ, CRZ and CTRZ (caused due to vortex breakdown), are clearly observed. At lower Re, the CTRZ is a weaker structure that exhibits a low-frequency unsteady flapping.

For the reacting flows, the mean axial velocity profiles are in good agreement with measurements, and slightly over-predicted close to the dump plane locations. This over-prediction is reflected by a more compact and attached flame in the predictions compared to the experimental observations which show a slightly lifted flame. Moreover, the predicted RMS fluctuations exhibit double peak in the burnt and un-burnt regions and on either side of the peak heat release.

The measured and predicted heat release distributions are in qualitative agreement with each other and exhibit the highest values along the inner edge of the shear layer. With increasing Reynolds number the flame region is seen to be more compact both experimentally and computationally.

This study demonstrates that the Thickened-Flame based LES approach with simplified chemistry for reacting flows is a promising tool to investigate reacting flows in complex geometries.

**ACKNOWLEDGMENTS**

This work was supported by the Clean Power and Energy Research Consortium (CPERC) of Louisiana through a grant from the Louisiana Board of Regents. Simulations are carried out on the computers provided by LONI network at Louisiana, USA (www.loni.org) and HPC resources at LSU, USA (www.hpc.lsu.edu). This support is gratefully acknowledged.

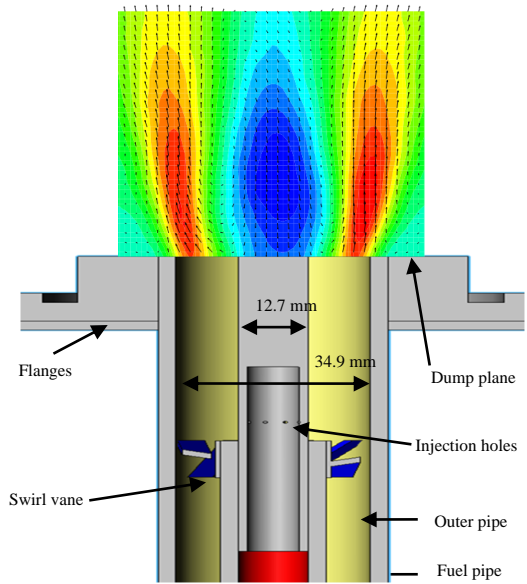

Figure 1. Sectional view of the swirl injector.

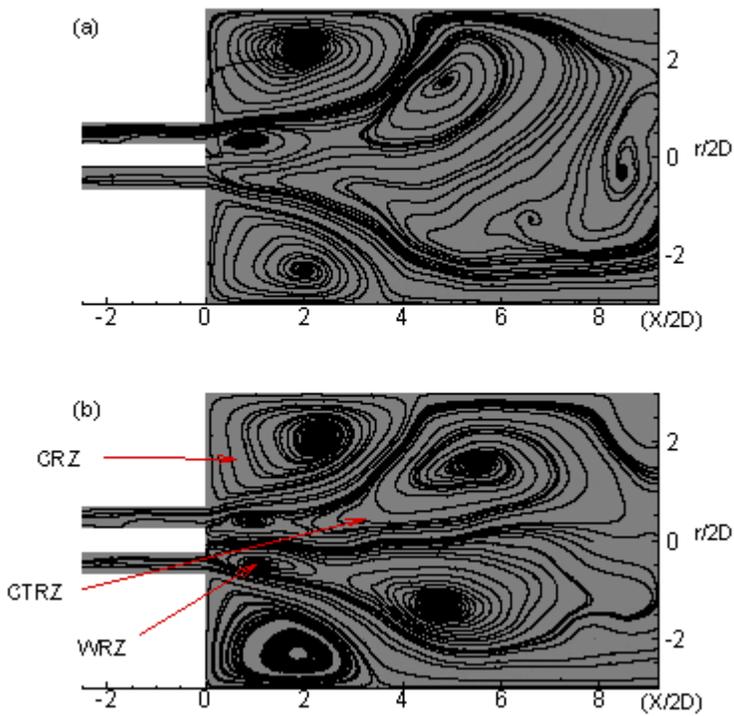

Figure 2. Streamline patterns for non-reacting flow condition [D: center-body diameter]: (a) Re=10144, (b) Re=13339

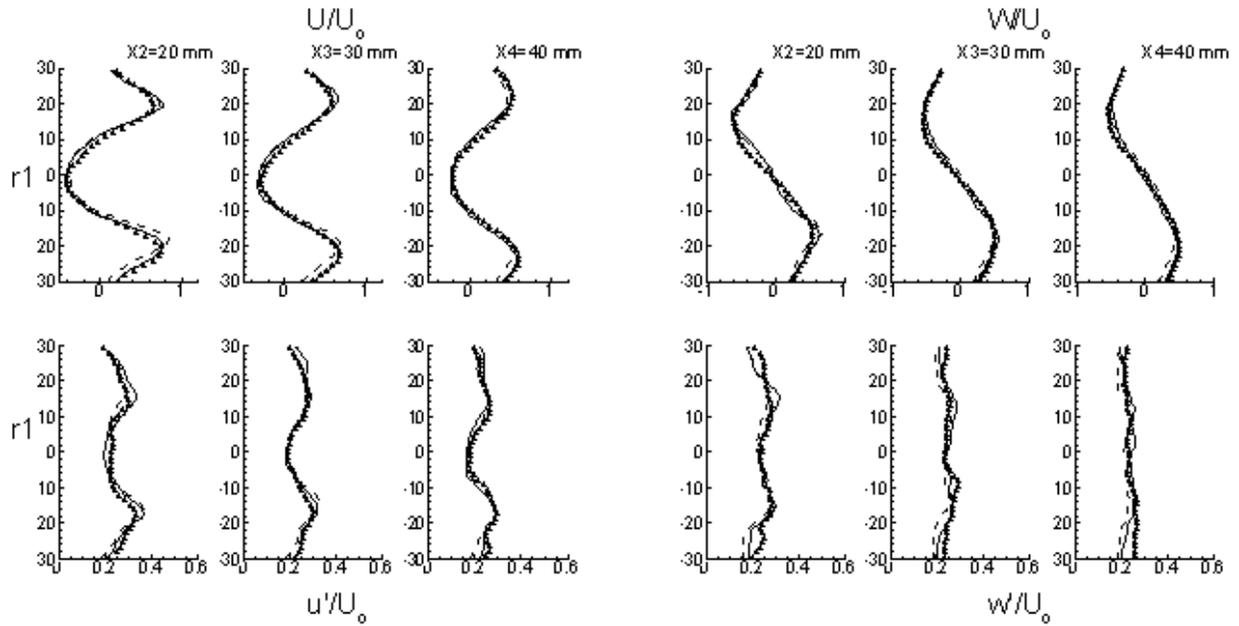

Figure 3. Non-reacting flow results for Re=13339 at different axial locations [r1=(r/2D) x25.4; X2, X3, X4=(X/2D) x25.4]: Experimental data (Δ), Lines are LES predictions: fine mesh ( —— ), coarse mesh ( ---- ). Mean axial velocity $U/U_o$, Mean tangential velocity $W/U_o$, Axial velocity fluctuation $u'/U_o$, Tangential velocity fluctuation $w'/U_o$.

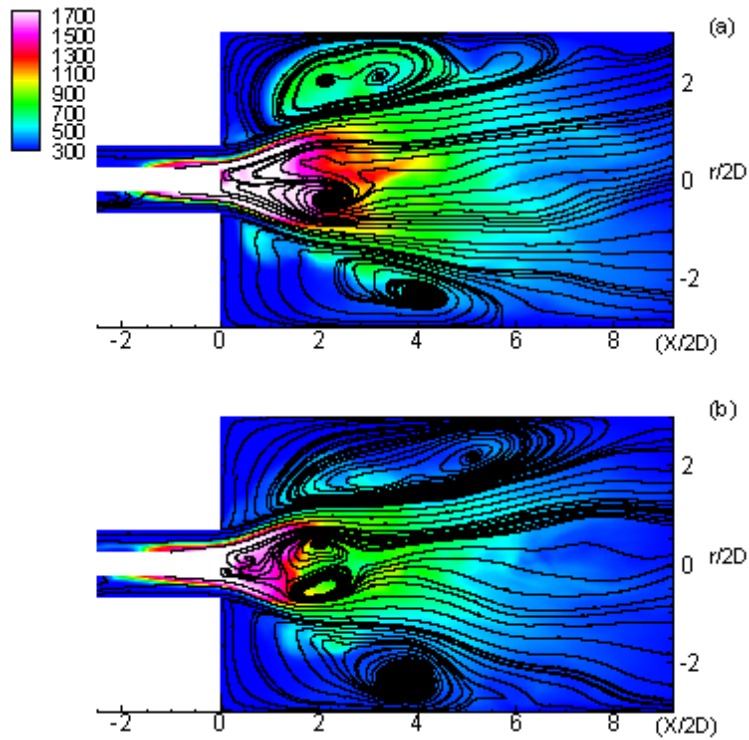

Figure 4. Mean temperature field streamline patterns for reacting flow condition: (a) Re=10144, (b) Re=13339

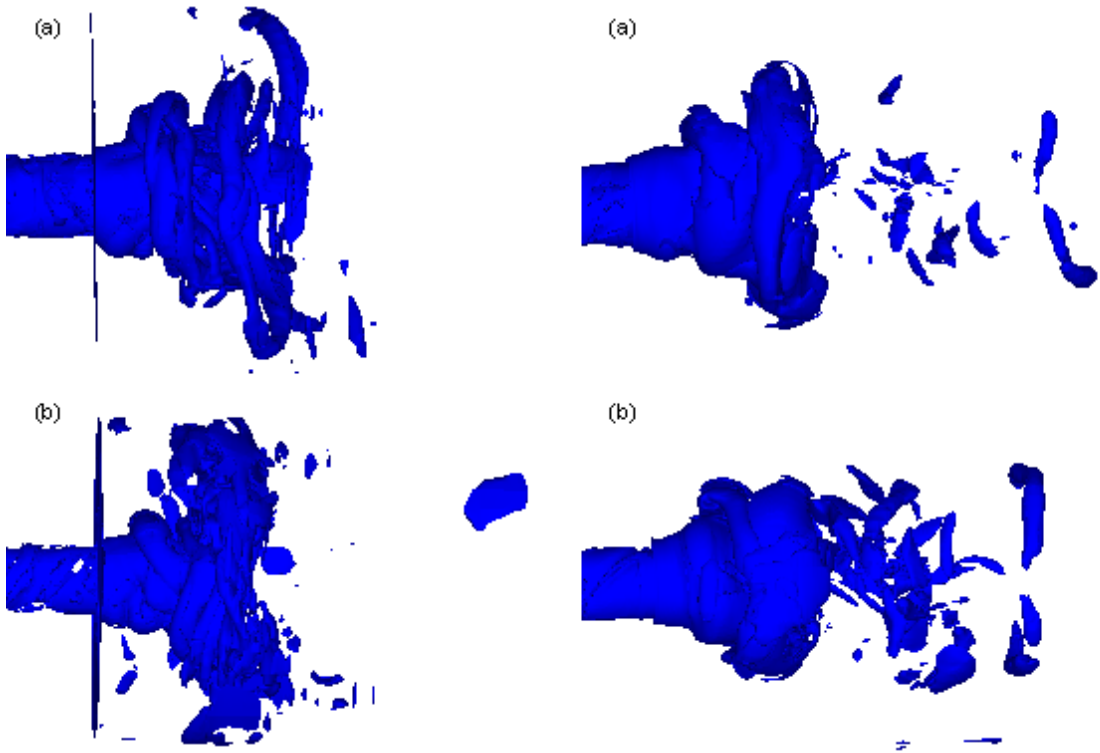

Figure 5.   Snapshots of iso-vorticity surface at $\omega=13$ s$^{-1}$ for non-reacting flow (left) and reacting flow (right) conditions: (a) Re=10144, (b) Re=13339

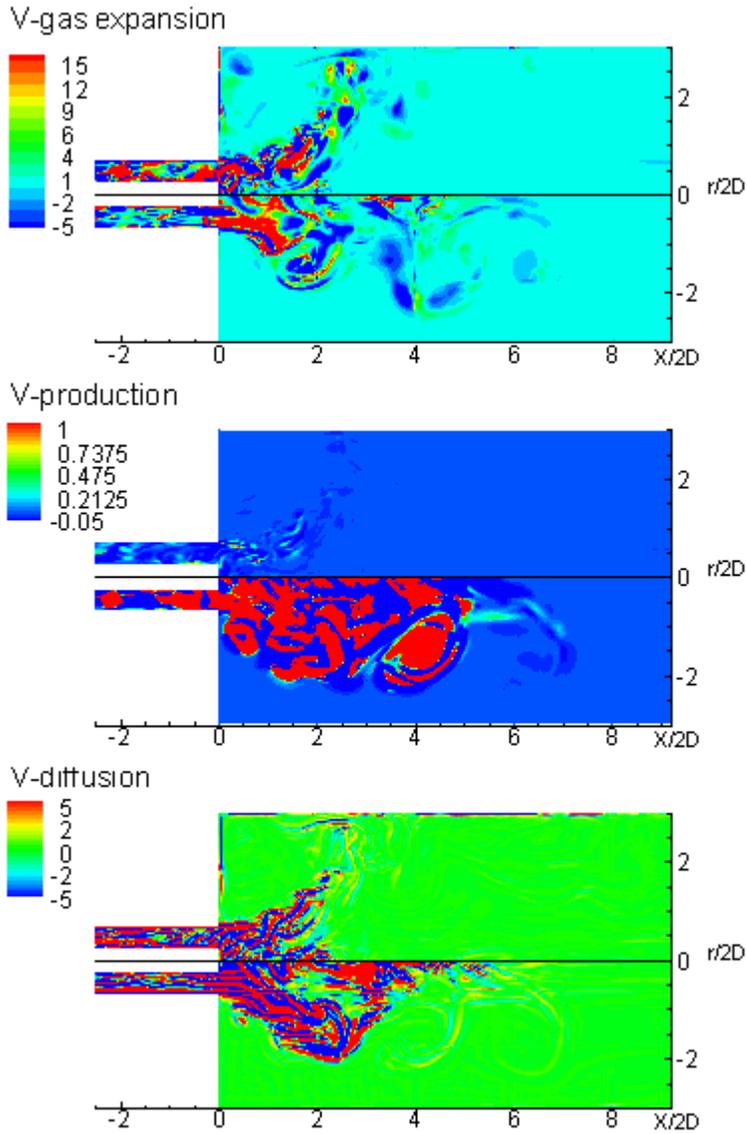

Figure 6. Snapshots of Gas expansion (top), baroclinic production (middle) and diffusion term (bottom) for non-reacting (upper half) and reacting flow (lower half) conditions for Re=13339

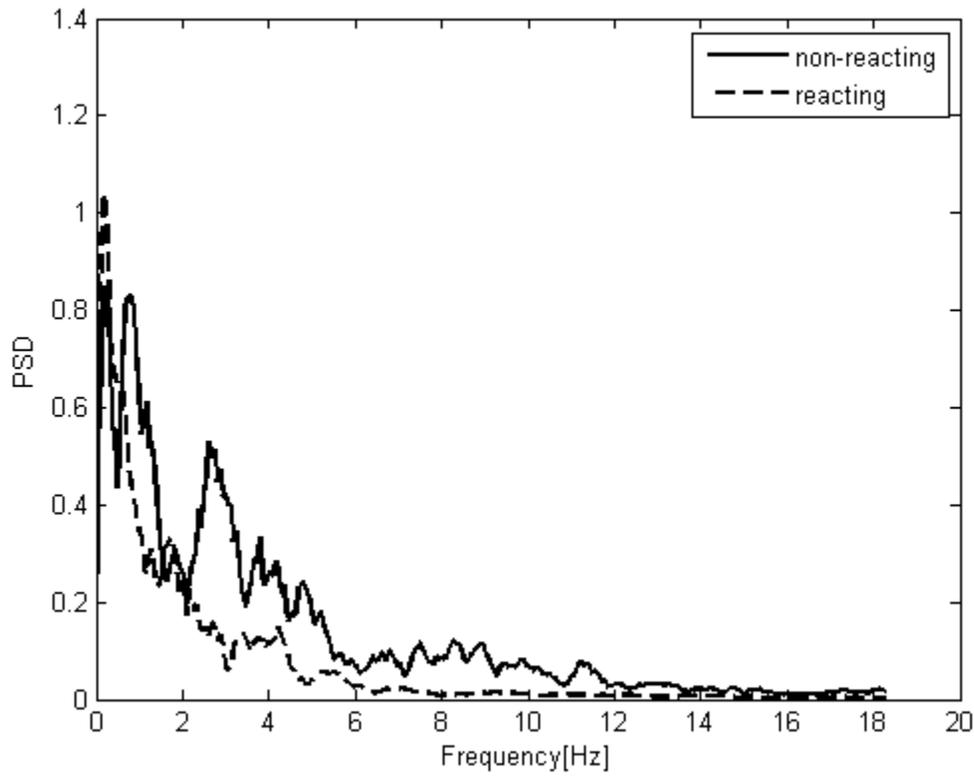

Figure 7.  Spectrum of axial velocity fluctuations for non reacting and reacting cases for Re=10144

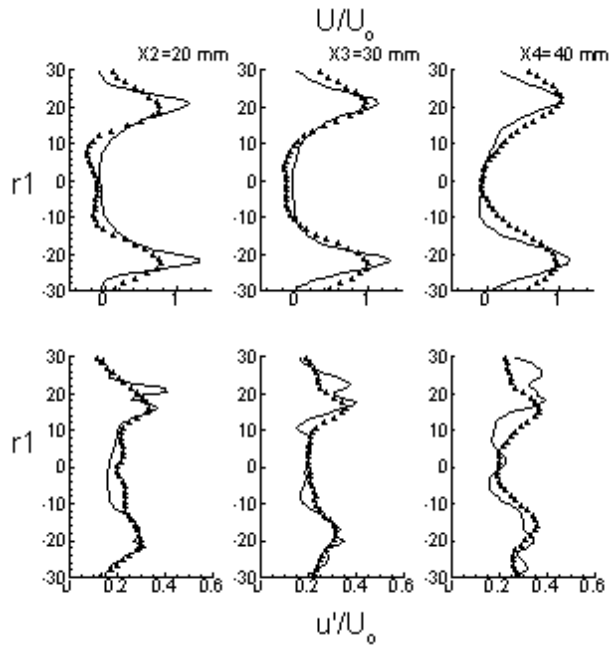

Figure 8. Reacting flow results for Re=13339 at different axial locations [r1=(r/2D) x25.4; X2, X3, X4=(X/2D) x25.4]: Experimental data (Δ), Lines are LES predictions ( — ). Axial velocity $U/U_o$, Axial velocity fluctuation $u'/U_o$

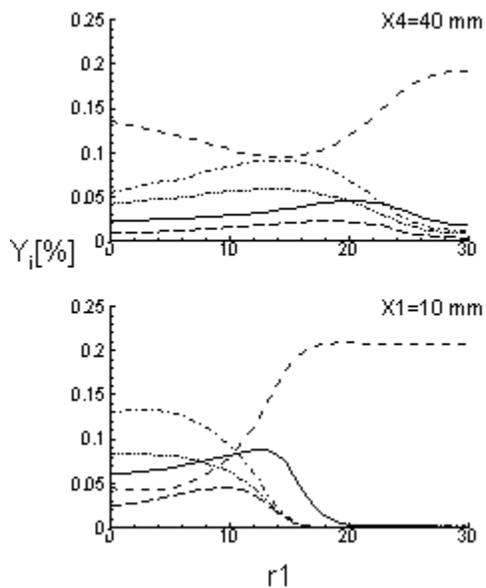

Figure 9. Mean species concentration for Re=13339 at two axial locations [r1=(r/2D) x25.4; X1, X4=(X/2D) x25.4]: $CH_4$ (—), $O_2$ (- - -), $H_2O$ (·· —), $CO_2$ (·-·-), $10\times CO$ (— —).

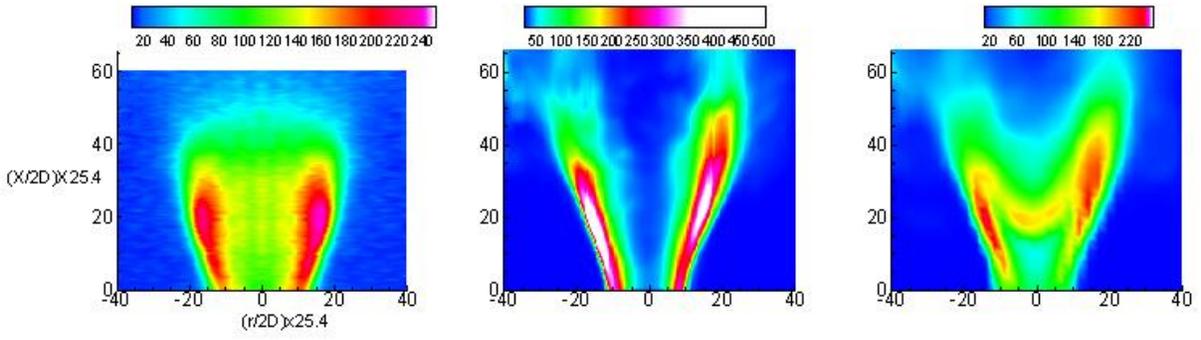

Figure 10. Experimental CH chemiluminescence measurement (left), Computational mean heat release (W/m$^3$) predictions at center plane (middle), Computational mean heat release (W/m$^3$) predictions averaged across flame (right) for Re=13339